\documentclass[referee]{aastex}
\graphicspath{{./pic/}}
\begin{document}

\title{Infall of nearby galaxies into the Virgo cluster as traced with HST\thanks{Based on observations made with the NASA/ESA Hubble Space Telescope,
  obtained at the Space Telescope Science Institute, which is
  operated by the Association of Universities for Research in Astronomy, Inc.,
  under NASA contract NAS 5-26555. These observations are
  associated with program GO 12878.}}

\author{Igor. D. Karachentsev}
\affil{Special Astrophysical Observatory RAS,  Nizhnij Arkhyz,    Karachai-Cherkessian Republic,
   Russia 369167}
\email{ikar@sao.ru}
\author{R. Brent Tully}
\affil{Institute for Astronomy, University of Hawaii, 2680 Woodlawn Drive, Honolulu, HI 96822, USA}

\author{Po-Feng Wu}
\affil{Institute for Astronomy, University of Hawaii, 2680 Woodlawn Drive, Honolulu, HI 96822, USA}

\author{Edward J. Shaya}
\affil{Department of Astronomy, University of Maryland, College Park, MD 20742, USA}

\author{Andrew E. Dolphin}
\affil{Raytheon Company, 1151 East Hermans Road, Tucson, AZ 85756, USA}

\begin{abstract}

  We measured the Tip of the Red Giant Branch distances to nine
galaxies in the direction to the Virgo cluster using the Advanced
Camera for Surveys on the Hubble Space Telescope. These distances
put seven galaxies: GR~34, UGC~7512, NGC~4517, IC~3583, NGC~4600, VCC~2037
and KDG~215 in front of the Virgo, and two galaxies: IC~3023, KDG~177
likely inside the cluster. Distances and radial velocities of the galaxies
situated between us and the Virgo core clearly exhibit the infall
phenomenon toward the cluster. In the case of spherically symmetric
radial infall we estimate the radius of the ``zero-velocity surface''
to be $(7.2\pm0.7)$ Mpc that yields the total mass of the Virgo cluster
to be $(8.0\pm2.3)\times10^{14} M_{\odot}$ in good agreement with its virial mass
estimates. We conclude that the Virgo outskirts does not contain
significant amounts of dark matter beyond its virial radius.

\end{abstract}

\maketitle

  \section{ Introduction}

  In the standard  $\Lambda$CDM cosmological model groups and clusters
are built from the merging of already formed galaxies embedded in massive
dark haloes (White \& Rees, 1978). Besides the dynamically evolved core, characterized by a
virial radius $R_v$, any cluster has a more extended region where galaxies
are falling towards the cluster center. In the simplest case of spherical
symmetry, the region of infall has a ``surface of zero-velocity'' at a radius
$R_0$ which separates the cluster against the global Hubble expansion.
The ratio $R_0/R_v$  lies in the range of (3.5 -- 4.0) being slightly
dependent on the adopted cosmological parameter $\Omega_{\Lambda}$ (Tully,
2010, Karachentsev, 2012).

As it has been noted by different authors (Vennik 1984, Tully 1987,
Crook et al. 2007, Makarov \& Karachentsev 2011, Karachentsev 2012), the total
virial masses of nearby groups and clusters leads to a mean local density of
matter of $\Omega_m\simeq0.08$, that is 1/3 the mean global density
$\Omega_m=0.24\pm0.03$ (Spergel et al. 2007). One possible explanation
of the disparity between the local and global density estimates may be
that the outskirts of groups and clusters contain significant amounts of
dark matter beyond their virial radii, beyond what is anticipated from
the integrated light of galaxies within the infall domain.  If so, to get
agreement between local and global values of $\Omega_m$, the total mass of
the Virgo cluster (and other clusters) must be 3 times their virial masses.
A measure of this  missing mass can be made by mapping the pattern of infall
into the cluster (or group). Uniquely in the case of the Virgo cluster, it
is possible to resolve the location of galaxies in three dimensions and
separate peculiar galaxies of infall from cosmic expansion as well as from
virial motions. The possibility of a massive dark superhalo around Virgo
can be easily tested using accurate distances at the near surface of the
Virgo infall boundary with Tip of the Red Giant Branch measurements.

 As shown by Lynden-Bell (1981) and Sandage (1986), in the case of a
spherical over density with cosmological parameter $\Lambda$ = 0 the radius
$R_0$ depends only on the total mass of a group (cluster) $M_T$ and the age
of the Universe $t_0$:

$$M_T=(\pi^2/8G) R_0^3 t^{-2}_0,                       \eqno(1) $$
where $G$ is the gravitational constant. Measuring $R_0$ via distances
and radial velocities of galaxies outside the virial radius of the system
$R_{v}$, one can determine the total mass of the system independent
of its virial mass estimate.

Numerous measurements of distances to nearby galaxies obtained recently
with the Hubble Space Telescope (HST) allowed us to investigate the
Hubble flow around the Local Group (Karachentsev et al. 2009) and some
other nearest groups: M~81 (Karachentsev \& Kashibadze, 2006), and Cen~A
(Karachentsev et al. 2006). The average total-to-virial mass ratio for the
proximate groups, derived from $R_0$ via eq.~(1) and from $R_{v}$, turns out
to be $<M_T/M_{v}> = 0.60\pm0.15$ (Karachentsev, 2005). But as it was noticed
by Peirani \& Pacheco (2006, 2008) and Karachentsev et al. (2007), in a flat
universe dominated by dark energy the resulting $M_T(R_0)$ mass is higher
than that derived from the canonical Lema\^\i{}tre-Tolman eq.~(1). In the
"concordant" cosmological model with $\Lambda$-term and $\Omega_m$ as a
matter component eq.~(1) takes a form

$$ M_T =(\pi^2/8G) R_0^3 H^2_0/f^2(\Omega_m), \eqno(2)$$

where

$$f(\Omega_m)=(1-\Omega_m)^{-1}-(\Omega_m/2)(1-\Omega_m)^{-3/2}
\arccos h[(2/\Omega_m)-1].\eqno(3)  $$

Assuming $\Omega_m=0.24$ and $H_0=72$ km s$^{-1}$ Mpc$^{-1}$, one can rewrite
(2) as

$$(M_T/M_{\sun}) = 2.12	\times10^{12}(R_0/Mpc)^3.         \eqno(4)$$

It yields a mass that is 1.5 as large as derived from the classic eq.~(1).
This correction leads to a good agreement on average between the $R_0$
mass estimates and virial masses for the  above mentioned galaxy groups.

  The most suitable object to explore the infall phenomena on a cluster
scale is the nearest
massive cluster of galaxies in Virgo. The kinematics and dynamics of
Virgo cluster infall were studied by Hoffman et al. (1980), Tonry \& Davis
(1981), Hoffman \& Salpeter (1982), Tully \& Shaya (1984), Teerikorpi et al.
(1992), and Ekholm et al. (1999, 2000). In a model developed by Tonry et al.
(2000, 2001) based on distance measurements of 300 E and S0 galaxies via
their surface brightness fluctuations, the Virgo cluster with its center
distance 17 Mpc and virial mass $M_v = 7\times10^{14}M_{\odot}$ generates an
infall velocity of the Local Group (LG) towards Virgo of about 140 km s$^{-1}$.
With this value of the virial mass, the expected radius of the infall zone
is $R_0$ = 7.0 Mpc or $\Theta_0$ = 23$^{\circ}$ in angular measure. Recently,
Karachentsev \& Nasonova (2010) considered the existing data on radial velocities
and distances of 454 galaxies situated within $\Theta$ = 30$^{\circ}$ around the
Virgo and came to the conclusion that the value of the radius $R_0$ lies in
the range [5.0 -- 7.5] Mpc. In the standard $\Lambda$CDM model with the
parameters $\Omega_m = 0.24$ and $H_0$ = 72 km s$^{-1}$ Mpc$^{-1}$
(Spergel et al. 2007), these quantities of $R_0$ correspond to a total
cluster mass $M_T = [2.7 - 8.9]	\times10^{14} M_{\odot}$.
The mass estimate derived from external galaxy motions does not contradict
the virial mass obtained from internal motions. However, the present accuracy
is insufficient to judge whether or not the periphery of the Virgo cluster
contains a significant amount of dark matter outside its virial radius
$R_v$ = 1.8 Mpc (Hoffmann et al. 1980).

  \section{ Expected pattern of the infall}

Fig. 1 represents the picture of Virgocentric infall based on current
observables collected by Karachentsev \& Nasonova (2010). It shows a relation
between radial velocities in the LG rest frame and distances of
galaxies within a cone of radius $\Theta_v = 6^{\circ}$, covering the
virialized core. Galaxy samples with distances derived by different
methods are marked by different symbols. The unperturbed
Hubble flow with a slope of $H_0$ = 72 km s$^{-1}$Mpc$^{-1}$ is given by
an inclined dashed line. The solid and dotted lines correspond to
the mean Hubble flow in a model of a point-like cluster mass with
$2.7\times10^{14}$ and $8.9\times10^{14} M_{\odot}$.

The distance of the Virgo cluster itself is now well established by
observations of Cepheid variables in 4 galaxies.
The Cepheid distances anchor precision {\it relative} distances for 84
galaxies with HST SBF measurements (Mei et al. 2007, Blakeslee et al. 2009)
and 4 galaxies with SNIa measurements (Jha et al. 2007).
These galaxies reside in the cluster core at $R_{LG} = (16.5\pm2)$ Mpc and
therefore are useless as tracers of the Virgocentric infall.

At large distances on the diagram, behind the Virgo cluster, while most
distance measures are based on the optical or IR Tully-Fisher relation with
typical errors of $\sim20$\%, there is one very well constrained group.
The Virgo W$^{\prime}$ group around NGC4365 (de Vaucouleurs 1961) with
$<V_{LG}> \simeq 1000$~km s$^{-1}$ contains one
galaxy with both a Cepheid and SNIa measurement and 5 other galaxies with
HST SBF measurements.  These observations locate Virgo W$^{\prime}$ at
23~Mpc, 6.5 Mpc behind Virgo.  The group velocity and distance
indicate that this group lies very near the edge of the Virgo infall zone
at $R_0$ on the far side of the cluster.

The most feasible way to trace the Z-like wave of Virgocentric infall in detail
is to make distance measurements to galaxies on the {\it front} side of the
cluster via TRGB. This method (Lee et al. 1993) is applicable to
galaxies of all morphological types and provides the needed distance
accuracy of $\sim$5--7\% (Rizzi et al. 2007).  The greatest precision will
be achieved with lines-of-sight tight to the cluster where projection factors
with radial motions will be minimal. Unfortunately, in the virial cone
$\Theta_v = 6^{\circ}$, there is no foreground galaxy with a literature TRGB distance.
In the wider area with $\Theta < 15^{\circ}$ there are only 2 galaxies:
NGC 4826 and GR-8 with existing TRGB distances between the LG and Virgo.

  \section{ Selection of targets}

  The scarcity of TRGB data on the near side of the Virgocentric infall
wave can be understood. In the past, targets for TRGB distance measurements
with HST were usually galaxies from the Kraan-Korteweg \& Tammann
(1979) sample with radial velocities $V_{LG}<500$ km s$^{-1}$. In the Virgo core
direction a galaxy with a velocity $\sim500$ km s$^{-1}$ may be a representative
of the Local Volume ($R_{LG}<10$ Mpc), or a Virgo cluster member,
or even be situated behind the cluster at $R_{LG}\simeq20$ Mpc and infalling
toward us. The selection of candidates that might be true nearby galaxies
hidden among the huge number of Virgo cluster members is a complicated task.
That is why Kraan-Korteweg \& Tammann (1979) even excluded the Virgo cluster
core ($\Theta < 6^{\circ}$) from their consideration.

  The expected number of missed nearby galaxies in the region $RA=[12.0^h -
13.0^h]$, and $Dec=[0^{\circ} - 25^{\circ}]$ can be estimated as a follows.
The Catalog of Neighboring Galaxies (Karachentsev et al. 2004) contains
450 objects with $R_{LG}<10$ Mpc distributed over the entire sky. In a
new version of the catalog by Karachentsev et al. 2013 (=UNGC), updated with
fresh data from recent optical and HI surveys (SDSS, HIPASS, ALFALFA, etc.)
there are about 800 candidates in almost the same volume to a radius
of 11 Mpc. Assuming that UNGC sample is $\sim 100$\% complete to $M_B = -12$ mag
and taking into account the inhomogeneous distribution of galaxies due
to the concentration towards the Supergalactic equator as well as the
presence of the Zone of Avoidance along the Milky Way, one can
estimate the expected number of nearby ($R_{LG}<11$) galaxies within the
identified $15^{\circ}\times25^{\circ}$ square as $\simeq 40$.

  We undertook a special search for likely foreground galaxies, inspecting
SDSS images of more than 2000 objects in the specified area. Among these we
found 37 galaxies with HI line widths that yield Tully-Fisher distances less
than $\sim$11 Mpc. Their radial velocities lie in
the range  $V_{LG}$ = ($ 400 - 1400)$ km s$^{-1}$, and the majority of
these turn out to be blue dwarf galaxies showing no apparent concentration
towards the Virgo center. As objects for our pilot program to measure
distances with ACS HST via TRGB, we selected 8 galaxies which have lower
Tully-Fisher distance estimates. In the target list we also included the S0-type
galaxy NGC 4600 with a distance estimate via surface brightness fluctuations
by Tonry et al. (2001). (The case of the nearby S0a galaxy NGC 4826 with D(sbf)= 7.48 Mpc
(Tonry et al. 2001) and D(trgb) = 4.37 Mpc (Jacobs et al. 2009) tells us
that these methods sometimes give distance estimates with a significant difference.
At present all nine our targets have been imaged with HST within GO 12878.

Galaxies situated on the
nearby boundary of the ``zero velocity sphere'' will have radial velocities
close to the mean cluster value, $<V_{Virgo}> = 1000$ km s$^{-1}$ and,
given the expected value $R_0 \simeq 7$ Mpc,
distances $R_{LG}\simeq10$ Mpc.
The F814W and F606W images of these galaxies obtained with ACS at HST in a
two orbit per object mode can determine their TRGB distances with an accuracy
of $\sim7$\% or $\sim$0.7 Mpc. Given a total mass of the cluster within
the radius $R_0$ expresses by eq.(4), then the measurement of $R_0\simeq 7$ Mpc
with an accuracy of $\sim 0.7$ Mpc can yield a mass of the Virgo cluster
with an error of $\sim30$\%.

  \section {Observations and data processing}

  We have observed 9 galaxies with the Advanced Camera
for Surveys (ACS) during the HST Cycle 20 (proposal 12878).
Between November 15, 2012 and March 30, 2013 we obtained
2080s F606W and 1640s F814W images of each galaxy using
ACS/WFC with exposures split to eliminate cosmic ray contamination.
The images were obtained from the STScI archive,
having been processed according to the standard ACS pipeline.
Stellar photometry was obtained using the ACS module of DOLPHOT
(http://americano.dolphinsim.com/dolphot), the successor to HSTPHOT
(Dolphin 2000), using the recommended recipe and parameters.  In brief,
this involves the following steps.  First, pixels that are flagged as bad
or saturated in the data quality images were marked in the data images.
Second, pixel area maps were applied to restore the correct count rates.
Finally, the photometry was run. In order to be reported, a star had to
be recovered with S/N of at least five in both filters, be relatively
clean of bad pixels (such that the DOLPHOT flags are zero) in both
filters, and pass our goodness of fit criteria ($\chi \le 2.5$ and
$\vert sharp \vert \le 0.3$). These restrictions reject non-stellar and blended objects.
At the high Galactic latitude of the Virgo cluster foreground
stars from the Milky Way are insignificant contaminants.
For some of the most distant galaxies we
extended to stars with $S/N > 2$ in order to evaluate the TRGB.
This extension introduces a lot of noise which is monitored by plotting CMD of
empty regions beside the galaxy body.

The TRGB is determined by a maximum
likelihood analysis monitored by recovery of artificial stars (Makarov
et al. 2006). Artificial stars with a wide range of known magnitudes and colors are
imposed at intervals over the surface of the target and recovered (or not) with the
standard analysis procedures
to determine both photometric errors and completeness in the crowded field environments.
The maximum likelihood procedure considers the luminosity function of stars
with colors consistent with the red giant branch after compensating for completeness
and assesses power law fits to the distributions above and below a break identified with the TRGB.
The slope of the power law faintward of the TRGB break is expected to be approximately
0.3 on a magnitude scale after correction for completeness.  If the RGB is sufficiently
observed to well below the tip then the slope can be a free parameter within a restricted range
but in the current cases with distances approaching the effective observational limits
the slope of the luminosity function fit below the TRGB is set to the expected value of 0.3.
Galactic extinction, minor at the polar location of the Virgo cluster,
is taken from Schlafly \& Finkbeiner
(2011).

The greatest potential for serious error with a TRGB measurement
comes about with confusion of the asymptotic giant branch (AGB) for the RGB.
Stars on the AGB that are burning both helium and hydrogen in shells
closely parallel and overlap the RGB on a CMD but rise as much as a magnitude
brighter.  Their peak brightness, dependent on age and metallicity, can be misinterpreted
as the TRGB.  AGB stars have intermediate ages of 1-10 Gyr although they are only
in sufficient quantity to be confusing at the lower end of that age range (Jacobs et al. 2011).
A general strategy that we employ is clipping of the area of the HST image
to avoid regions of young and intermediate age stars (and regions beyond the target
dominated by background and foreground contaminants) in order to maximize the contrast
of the old population contributing to the RGB.

The calibration of the absolute value of the TRGB including
a small color term has been described by Rizzi et al. (2007).
The RGB is redder for older or more metal rich populations but
galaxies inevitably have old and metal poor components, resulting in
reasonable stability of tip magnitudes in the F814W band.
Images, color-magnitude diagrams, photometry tables, TRGB measurements, and distance determinations are
made available at http://edd.ifa.hawaii.edu by selecting the catalog
CMDs/TRGB (Jacobs et al 2009).

  \section{ TRGB distances to nine target galaxies}

 Images of our target galaxies taken from Sloan Digital Sky Survey
(http://www.sdss.org/) are shown in Figure 2. Each field has a size
of 6 by 6 arcminutes. North is up and East is left.
The ACS HST footprints are superimposed on the SDSS frames.
In Figure 3 a mosaic of enlarged ACS (F606W + F814W) images of
the nine galaxies are shown. Their size is 1 arcminute, North is up
and East is left. Color magnitude diagrams (CMDs) of F814W versus
(F606W - F814W) are presented in Figure 4.

A summary of some basic parameters for the observed galaxies as well as
the resulting distance moduli for them are given in Table~1.
Some additional comments about the galaxy properties are briefly
discussed below.

\begin{table}[ht!]\footnotesize
\caption{Target galaxies in front of the Virgo cluster observed with HST.}
\begin{tabular}{lcrrrlrrrccc} \hline
 Name   &  RA (J2000) Dec & $V_{LG}$& D   & $\Theta$ & $B_T$ &  T  &
$m_{FUV}$&$m_{21}$& $W_{50}$& $I_{TRGB}$ &$D_{HST}$ \\
\hline
 (1)    &         (2)     &    (3)  &   (4)   &  (5)       & (6 )  & (7) & (8)
   & (9)   & (10)   & (11) & (12)    \\
\hline
IC3023  &  121001.7+142201&   710   & 7.7 tf  &  5.4      & 15.35 & 10  &16.78
  & 16.33 & 44     & $\sim27$              & $\sim 17$       \\
GR34    &  122207.6+154757&  1205   & 8.9 tf  &  4.0      & 15.95 & 10  &19.04
  & 18.24 & 25     & 25.94$^{+.28}_{-.15}$ & 9.29$\pm$0.93     \\
U7512   &  122541.3+020932&  1354   &10.6 tf  & 10.3      & 15.20 & 10  &17.40
  & 15.31 & 65     & 26.37$^{+.08}_{-.09}$ & 11.8$\pm$1.2       \\
N4517   &  123245.5+000654&   978   & 9.7 tf  & 12.3      & 11.09 &  7  &15.86
  & 12.39 &307     & 25.67$^{+.16}_{-.12}$ & 8.34$\pm$0.83     \\
IC3583  &  123643.5+131534&  1024   & 7.6 tf  &  1.7      & 13.31 &  9  &15.43
  & 15.66 &105     & 26.04$^{+.06}_{-.05}$ & 9.52$\pm$0.95      \\
KDG177  &  123958.5+134653&   913   & 8.2 tf  &  2.6      & 16.36 & 10  &18.47
  & 16.17 & 30     & $\sim27$                 & $\sim 17$       \\
N4600   &  124023.0+030704&   713   & 7.4 sb &  9.6      & 13.70 &  0  &20.38
  &   --- & ---    & 25.78$^{+.05}_{-.05}$ & 8.90$\pm$0.89      \\
VCC2037 & 124615.3+101212 &  1038    &7.4 tf   &  3.3      & 15.80 & 10  &17.90
  & 18.50 & 29     & 26.01$^{+.22}_{-.17}$ & 9.63$\pm$0.96      \\
KDG215  &  125540.5+191233&   362   & 5.5 tf  &  9.1      & 16.90 & 10  &18.86
  & 15.79 & 25     & 24.33$^{+.07}_{-.06}$ & 4.83$\pm$0.34      \\ \hline
 \end{tabular}
 \end{table}

(1) galaxy name, (2) equatorial coordinates, (3) radial velocity in
km s$^{-1}$ in the LG rest frame from NASA Extragalactic Database
(http://ned.ipac.caltech.edu/), (4) linear distance (in Mpc) as given in
UNGC, estimated via Tully-Fisher relation (tf) or from surface
brightness fluctuations (sb), (5) angular separation $\Theta$ (in degrees)
from the Virgo cluster center that has been identified with NGC4486, (6) apparent
integrated B- magnitude as given in UNGC, (7) morphological type in de
Vaucouleurs scale, (8) far-ultraviolet integrated magnitude from the Galaxy
Evolution Explorer (GALEX) space telescope (Gil de Paz et al. 2007), (9) HI-line
magnitude $m_{21} = 17.4 - 2.5 \log F_{HI}$, where $F_{HI}$ is an HI-flux in
Jy km/s from Haynes et al (2011) or the Lyon Extragalactic Database =LEDA
(http://leda.univ-lyon1.fr/),(10) HI-line width (in km/s) at the 50\%
level of the maximum, (11) TRGB magnitude and its 68\% uncertainty from the
maximum likelihood analysis.
(12) the linear distance (in Mpc) and conservative global characterization of
10\% uncertainty for a one-orbit ACS observation of a galaxy near 10 Mpc.


 {\bf GR34=VCC530, UGC7512 and VCC2037.} These are irregular type dwarf galaxies with
narrow HI lines. New TRGB distances to them agree with the
Tully-Fisher distances confirming all the galaxies to be situated in
front of Virgo cluster.

 {\bf NGC4517.} This Sd galaxy seen edge-on has the major angular diameter
about 12', extending far beyond the ACS frame. Its CMD is constructed from an
outskirts field along the minor axis to sample the halo and avoid crowded dusty regions of star
formation. The TF distance to NGC~4517 is consistent with
the TRGB distance.

\suppressfloats

 {\bf IC3583.} This Magellanic type dwarf has an asymmetric diffuse halo
extended to the West.  The field contributing to the CMD that is shown in Fig. 4
is clipped to minimize young and intermediate age populations
and optimize the contribution of the old population.  See Figure 5 for the CMD
for the full ACS field
and an outline of the excised region containing many young stars.
Together with a bright spiral galaxy NGC~4569, IC3583 forms the
optical pair Arp 76 having a radial velocity difference of 1245 km/s.
The NW part of NGC~4569 is seen in the SE corner of the ACS frame.
Our estimate of distance to NGC~4569 via its TRGB yields $D > 17$ Mpc.

 {\bf NGC4600.} This is a gas-poor dwarf lenticular galaxy with $H\alpha$
emission in the core (Karachentsev \& Kaisin, 2010).  We recognize a moderate
agreement in the distance estimates for NGC~4600 via surface brightness
fluctuations (Tonry et al. 2001) and from TRGB. It is a bit unexpected
to find this isolated dS0 galaxy in front of the Virgo cluster rather than in the virial
core.

 {\bf KDG215=LEDA44055.} This galaxy is gas-rich low surface brightness dwarf with a
narrow HI-line, a high hydrogen mass-to-stellar mass ratio $M_{HI}/M_* = 3.1$,
and a narrow RGB characteristic of a low metallicity system.  KDG215 lies more than
a magnitude closer than any of the other targets, at 4.8 Mpc.

 {\bf IC3023 and KDG177=VCC1816.} Both the galaxies of Im- type are HI-rich
and active star formation objects typical of field galaxies.
In spite of their narrow HI-lines: 44 and 30 km/s, they both appear to belong to
the Virgo cluster.   The TRGB are not seen as would have been the case if these
galaxies were in the Virgo foreground.  In each case, the TRGB is probably being seen
around $I \sim 27$ as expected for a cluster member.  These tentative measurements are
at the limit of the current HST photometry and we do not attempt a distance determination.

  Apart from these objects, there are five other galaxies in front of the Virgo
cluster that have accurate distance measurements. Information about them is collected in Table 2.
We use the data on distances and radial velocities of these 7 + 5 galaxies from
Tables 1 and 2 to trace the near-side Virgocentric infall. Two probable Virgo core galaxies with
uncertain distances: IC 3023 and KDG 177 are excluded from consideration.
In addition, the analysis will include the galaxy NGC 4365 in the Virgo W' cloud
as a representative with an accurate distance of the back-side infall to the Virgo cluster.
Its parameters are given in the last line of Table 2.

\begin{table}[th!]
\caption{Other galaxies in front or back of the Virgo cluster with accurate distance
measurements}
\begin{tabular}{lcrrlrrrl} \hline
 Name & RA (J2000) Dec  & $V_{LG}$  &  D  &    & $\Theta$& $B_T$ &  T &
Reference
         \\
\hline
N4527 & 123408.4+023913 & 1591  &14.1$\pm$1.4 & SN  & 9.8  &11.38&  4 & Jha et al. 2007
   \\
N4536 & 123427.0+021117 & 1662  &14.3$\pm$1.4 & cep &10.3  &11.16&  4 & Riess et al. 2005
   \\
N4725 & 125026.6+253003 & 1176  &12.4$\pm$1.2 & cep &13.9  &10.11&  2 & Freedman et
al. 2001
\\
N4826 & 125644.2+214105 &  365  & 4.37$\pm$0.44& TRGB &11.2  & 9.30&  2 & Jacobs et
al. 2009
   \\
GR-8  & 125840.4+141303 &  139  & 2.13$\pm$0.21& TRGB & 9.1  &14.79& 10 & Tully et al. 2006
    \\
\hline
N4365 & 122428.3+071904 & 1112  &23.1$\pm$2.3 & sbf & 5.3  &10.52& -3 & Blakeslee et
al. 2009
 \\
\hline
\multicolumn{9}{p{\textwidth}}{
The column designations are similar to those in Table 1.
}
\end{tabular}
\end{table}

 \section {Estimating the total mass of the Virgo cluster}
 As noted above, the analysis of available observational data
on radial velocities and distances for several hundred galaxies in the vicinity
of the Virgo cluster lead to the conclusion that the radius of the zero velocity
surface of the cluster lies in the range $R_0 = (5.0 - 7.5)$ Mpc (Karachentsev \& Nasonova, 2010).
According to equation (4) this scatter in $R_0$ leads to a wide scatter in the total mass
estimates of the cluster, $M_T = (2.7 - 8.9)\times10^{14} M_{\odot}$, exceeding a factor
of three. New accurate distance measurements to relatively few galaxies
residing near the front side of Virgo fix the $R_0$ and $M_T$ quantities
in a narrower interval.

Figure 6 reproduces a pattern of the Hubble flow in front and back of the Virgo cluster
restricted to the most accurate constraints.
Compared with Figure 1, it exhibits a much more distinct character of the
infall. Open circles in the Fig. 6 show the galaxies from Table 2 with
accurate distance estimates. The solid circles correspond to seven galaxies
in front of the Virgo cluster with distances measured in this program with HST. The
horizontal bars indicate  distance errors. The grey vertical column
denotes the zone of virial motions corresponding to the mean distance to
the cluster  $\langle D\rangle = 16.5\pm0.4$ Mpc (Mei et al. 2007) and the virial radius
$R_v = 1.8$ Mpc. The inclined dashed line indicates the unperturbed Hubble
flow with the parameter $H_0 = 72$ km/s/Mpc.

  The average angular separation of the 12 galaxies situated in front of
Virgo from its center is $\langle\Theta_v\rangle = 8.7^{\circ}$. The solid wave-like line
in the figure reproduces the behaviour of Hubble flow
perturbed by a point-like mass  $M_T = 8.0\times 10^{14} M_{\odot}$ at the average
angular separation $8.7^{\circ}$ from the Virgo center.

  To determine the radius $R_0$, one needs to fix the mean radial velocity
of the cluster, $\langle V_{Virgo}\rangle_{LG}$  in the rest frame of the Local Group.
According to Binggeli et al. (1993), it equals to $+946\pm35$ km/s. This
estimate was obtained over a large number of galaxies with measured velocities
but unmeasured distances, whose membership in the Virgo cluster was considered
to be probable. Basing on the galaxies with membership in Virgo
confirmed by accurate distances, Mei et al. (2007) derived the mean
cluster velocity of $+1004\pm70$ km/s. The difference of 58 km/s between
these estimates can be caused by a specific selection affecting Binggeli's
estimate. In a spherical layer between the radii $R_v$ and $R_0$ bounded by a
cone with the angular radius of $\Theta_0 \sim 20^{\circ}$, the expected number of
galaxies behind the cluster is greater than that in front of the cluster.
In the case of radially infalling galaxies into the cluster core, the
difference in galaxy number falling toward us and away from us should artificially
decrease the mean radial velocity of the sample. Probably, any (unknown)
pre-selection effect on velocities could also be in the list of targets
investigated by Mei et al. (2007).  We adopt the average of these two
independent values as the radial velocity of the Virgo cluster centroid,
$\langle V_{Virgo}\rangle_{LG} = 975\pm29$ km/s, shown in Figure 6 as
the horizontal dashed line.

  As seen from Fig. 6, the straight line of unperturbed Hubble flow with
the parameter $H_0 = 72$ km/s/Mpc crosses the Virgo center at
$V_{LG} = +1188$ km/s which corresponds to the infall velocity of LG toward
the Virgo: $\Delta V_{LG} = (975\pm29) - 72 (16.5\pm0.4) = - 213\pm41$ km/s.
This quantity is not significantly higher than the previous estimates --139 km/s (Tonry
et al. 2001) and --185 km/s (Tully et al. 2008).

  The presented data exhibit also that the solid wave-like line crosses
the line of the mean cluster velocity at the distance of 9.3 Mpc. Therefore,
the radius of the zero-velocity surface around the Virgo cluster turns out to be
$R_0 = 16.5 - 9.3 = 7.2$ Mpc. There are at least
three circumstances affecting this estimate: a) uncertainty of the Virgo
center position, which is $\sim0.4$ Mpc, b) uncertainty of the mean velocity of
the cluster $\sim$30 km/s corresponding to $\sim0.3$ Mpc on the distance scale, and \\
c) the mean-square scatter of galaxies with respect to the Z-like line that
consists of $\sim0.5$ Mpc. Considering these factors as being statistically
independent, we obtain the sought-for radius

	       $$R_0 = (7.2\pm0.7) {\rm Mpc}.$$

According to equation (4), this quantity corresponds to the total mass
of the Virgo cluster

	       $$M_T = (8.0\pm2.3)\times 10^{14} M_{\odot}.$$

Virial mass estimates for the Virgo are: 6.2 (de Vaucouleurs, 1960),
7.5 (Tully \& Shaya, 1984) and 7.2 (Giraud, 1999) in units of $10^{14} M_{\odot}$.
These values all have been normalized to the Virgo cluster distance of 16.5 Mpc.
As one can see, the total cluster mass estimate via $R_0$ is consistent
with the average virial mass estimate, $M_v = (7.0\pm0.4)\times 10^{14} M_{\odot}$.
Consequently, the zone of infall, at a radius 4 times the virial
radius (assuming $\Omega_m = 0.24$), does not contain a large amount of
mass outside $R_v$. This conclusion agrees with the results of N-body
simulations performed by Rines \& Diaferio (2006) and Anderhalden \& Diemand
(2011) for a cluster dark matter profile. These authors obtained the
$M_T/M_v$ ratio to be 1.19 and 1.25, respectively.

  We draw attention to the regularity of the infall pattern seen in front of
the Virgo cluster. A scatter of 12 galaxies along the vertical scale with respect
to the Z-shape line under parameters  $M_T = 8.0\times 10^{14} M_{\odot}$ and
$\langle \Theta\rangle = 8.7^{\circ}$ corresponds to $\sigma_v = 155$ km/s. When the
difference of the individual $\Theta$ of the galaxies is taken into account,
the value of $\sigma_v$ drops to 130 km/s. An essential part of this scatter,
 $\sim90$ km/s, is caused by errors of the distance measurements, which are $\sim(7-10)$\%.
 After a quadratic subtraction of the
component related to distance errors, the remaining (``cosmic'') dispersion
of radial velocities turns out to be $\sim95$ km/s. Therefore, one can say that
the infall flow pattern around the Virgo cluster looks to be rather ``cold''.

	     \section{Concluding remarks.}

 The measurements of distances to nearby galaxies with the Hubble Space
Telescope makes the picture of galaxy infall into the Virgo cluster much more
distinct. Among nine galaxies selected as Virgo foreground candidates
for our pilot HST GO 12878 program, seven reside in the expected near
region  while two others are probably cluster members. In our list of
targets for HST there are $\sim30$ more galaxies with Tully-Fisher
distances around 10 Mpc. Measurements of their distances with ACS HST
can give us a more precise estimate of the total mass of the nearest large
cluster via infalling galaxy motions.
Multicolor images of galaxies that have been obtained
with the 3.5-meter CFHT telescope under the program ``Next Generation Virgo
Cluster Survey'' (Ferrarese et al. 2012) will be useful in choosing the
best candidates for new HST observations.

 In the framework of the simplest spherically-symmetric radial infall
of galaxies into a point-like central mass, the observed distances and
radial velocities of galaxies in front of Virgo yield the value of total
mass of the cluster in good agreement with the virial mass:  $M_T =
(1.14\pm0.35) M_v$. It should be stressed here that the quantity
$M_T = (8.0\pm2.3)\times 10^{14} M_{\odot}$ was obtained in the case of standard $\Lambda$CDM
model with the parameter $\Omega_{\Lambda} = 0.76$. In the old cosmological
model with $\Omega_{\Lambda} = 0$, the estimate of total mass of the Virgo cluster via
motions of surrounding galaxies would be (35--40)\% lower, pushing mass estimates
almost out of the confidence interval below virial mass estimates derived
via internal motions. This circumstance can be considered as another
display of the existence of dark energy on a local scale of $\sim10$ Mpc.
It can be noted that Tully \& Shaya (1984) had already used a similar argument to suggest that
a $\Lambda$- term might be appropriate to explain the total kinematic pattern of the Virgo cluster.

  According to our estimate, the Hubble flow around the Virgo cluster
looks to be rather cold with a characteristic line-of-sight scatter
$\sim95$ km/s. This preliminary result, if confirmed, may impose constraints
on some models of cluster formation.  More new accurate distance
measurements with HST are required to check this claim.

  As was noted by Karachentsev et al. (2003) and Tully et al. (2008, 2013),
the nearby galaxies residing inside a radius of $\sim6$ Mpc around the Local
Group form a flat configuration (the ``Local Sheet'') with
surprisingly low peculiar velocities of the barycenters of groups of $\sim30$ km/s,
A hint to the existence of the Local Sheet can be seen in Figure 6 too, where
three the nearest galaxies: GR-8, NGC~4826, and KDG~215  all within D = 5 Mpc,
follow remarkably well the unperturbed Hubble flow. To our knowledge,
the existence of such calm domain structures, like the Local Sheet, still has
not sufficiently attracted the attention of cosmologists.

 {\bf Acknowledgements}.
The authors thank anonymous referee for thorough examination of the manuscript
and for useful comments and suggestions to improve the text.
Support for the program GO 12878 was provided by NASA through a grant from the Space
Telescope Science Institute, which is operated by the Association of Universities for
Research in Astronomy, Inc., under NASA contract NAS 5-26555.
I.K. acknowledges support by RFBR-DST grant 13-02-92690 and RFBR-DFG grant 12-02-91338.

\clearpage
{\bf References.}

Anderhalden D., Diemand J., 2011, MNRAS, 414, 3166

Binggeli B., Popescu C.C., Tammann G.A., 1993, A\&A, 98, 275

Blakeslee, J.P., Jordan, A., Mei, S., et al. 2009, ApJ, 694, 556

Crook A.C., Huchra J.P., Martimbeau N. et al., 2007, ApJ, 655, 790

de Vaucouleurs, G. 1961, ApJS, 6, 213

de Vaucouleurs, G. 1960, ApJ, 131, 585

Dolphin, A. 2000, PASP, 112, 1383

Ekholm T., Lanoix P., Teerikorpi P., Fouque P., Paturel G., 2000, A\&A, 355,835

Ekholm T., Lanoix P., Teerikorpi P., Paturel G., Fouque P., 1999, A\&A, 351, 827

Ferrarese L., Cote P., Cuillandre J. et al. 2012, ApJS, 200, 4

Freedmann W.L., Madore B.F., Gibson B.K. et al. 2001, ApJ, 553, 47

Gil de Paz A., Boissier S., Madore B.F., et al. 2007, ApJS, 173, 185

Giraud E., 1999, ApJ, 524, L15

Haynes M.P., Giovanelli R., Martin A.M., et al. 2011, AJ, 142, 170

Hoffman G.L., Salpeter E.E., 1982, ApJ, 263, 485

Hoffman G.L., Olson D.W., Salpeter E.E., 1980, ApJ, 242, 861

Jha, S, Riess, A.G., Kirshner, R.P. 2007, ApJ, 659, 122

Jacobs, B.A., Rizzi, L, Tully, R.B. et al. 2009, AJ, 138, 332

Jacobs, B.A., Tully, R.B., Rizzi, L., et al. 2011, AJ, 141, 106

Karachentsev I.D., Makarov D.I., Kaisina E.I., 2013, AJ, 145, 101 (UNGC)

Karachentsev I.D., 2012, Ast.Bull. 67, 115

Karachentsev I.D., Nasonova O.G., 2010, MNRAS, 405, 1075

Karachentsev I.D., Kaisin S.S., 2010, AJ, 140, 1241

Karachentsev I.D., 2005, AJ, 129, 178

Karachentsev I.D., Karachentseva V.E., Huchtmeier W.K., Makarov D.I.,
   2004, AJ, 127, 2031 (CNG)

Karachentsev I.D., Makarov D.I., Sharina M.E., et al. 2003, A\&A, 398, 479

Kraan-Korteweg R.C., Tammann G.A., 1979, Astron. Nachr., 300, 181

Lee M.G., Freedman W.L., Madore B.F., 1993, AJ, 106, 964

Lynden-Bell D., 1981, The Observatory, 101, 111

Makarov D.I., Karachentsev I.D., 2011, MNRAS, 412, 2498

Makarov, D.I, Makarova, L., Rizzi, L. et al. 2006, AJ, 132, 2729

Mei S., Blakeslee J.P., C\^ot\'e P., Tonry J.L., et al., 2007, ApJ, 655, 144

Peirani S, Pacheco J.A., 2008, A \& A, 488, 845

Peirani S, Pacheco J.A., 2006, New Astr., 11, 325

Riess A.G., Li W., Stetson P.B., et al. 2005, ApJ, 627, 579

Rines K., Diaferio A., 2006, AJ, 132, 1275

Rizzi L., Tully R.B., Makarov D.I., et al., 2007, ApJ, 661, 813

Sandage A., 1986, ApJ, 307, 1

Schlafly, E.F., Finkbeiner, D.P., 2011, ApJ, 737, 103

Spergel D.N., et al., 2007, ApJS, 170, 377

Teerikorpi P., Bottinelli L., Gouguenheim L., Paturel G., 1992, A\&A, 260, 17

Tonry J.L., Schmidt B.P., Barris B., et al., 2003, ApJ, 594, 1

Tonry J.L., Dressler A., Blakeslee J.P., et al., 2001, ApJ, 546, 681

Tonry J.L., Davis M., 1981, ApJ, 246, 680

Tully R.B., Courtois H.M., Dolphin A.E., et al. 2013, AJ, 146, 86

Tully R.B., Shaya E.J., Karachentsev I.D., et al., 2008, ApJ, 676, 184

Tully R.B., Rizzi L., Dolphin A.E., et al., 2006, AJ, 132, 729

Tully R.B., 1987, ApJ, 321, 280

Tully R.B., Shaya E.J., 1984, ApJ, 281, 31

Tully R.B., 2010, arXiv:1010.3787

Vennik J., 1984, Tartu Astron. Obs. Publ., 73, 1

White S.D.M., Rees M.J., 1978, MNRAS, 183, 341

\begin{figure}[th!]
\includegraphics[scale=0.6,angle=-90]{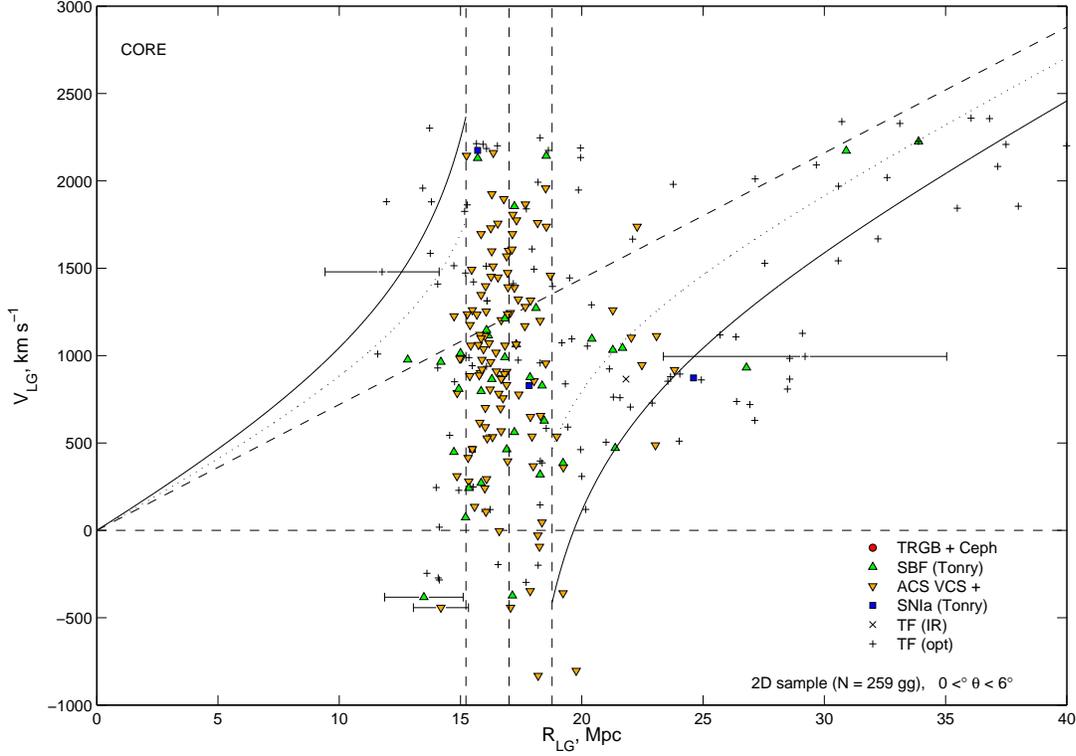}
 \caption{The radial velocity vs. distance relation for galaxies in the Virgo
cluster region with respect to the Local Group centroid, as shown in
Fig. 1 by Karachentsev \& Nasonova (2010). Galaxy samples with distances
derived by different methods are marked by different symbols. The inclined
line traces the unperturbed Hubble relation with the global Hubble parameter
$H_0 = 72$ km s$^{-1}$ Mpc$^{-1}$. The vertical dashed lines outline the virial
zone.
The solid and dotted lines correspond to Hubble flow perturbed by virial masses
of $2.7	\times 10^{14}$ and $8.9\times 10^{14} M_{\odot}$ as the limiting cases
within the confidence
range at $\Theta = 0^{\circ}$. The typical distance error bars for each dataset
are
shown.}
\end{figure}
\clearpage

\begin{figure}
\includegraphics[scale=0.3]{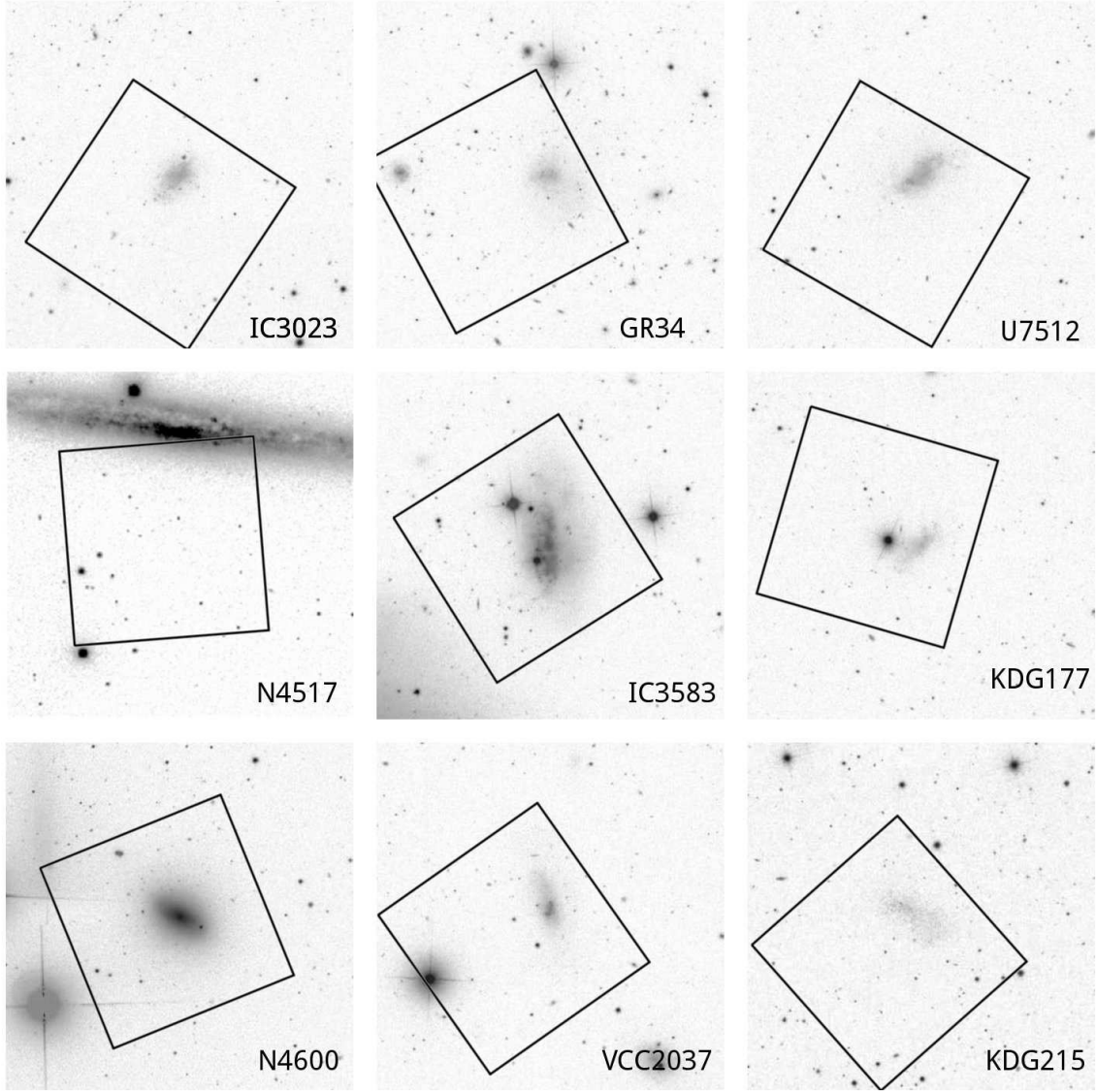}
\caption{Sloan Digital Sky Survey images of nine target galaxies.
Each field has a size of 6 by 6 arcminutes. North is up and East is left.
The HST ACS footprints are superimposed.}
\end{figure}

\begin{figure}
\def\FIG#1#2{\includegraphics[width=0.32\textwidth]{#1}\par\noindent\centerline{
\sf\Large{#2}}}
\begin{tabular}{p{0.33\textwidth}p{0.33\textwidth}p{0.33\textwidth}}\\
\FIG{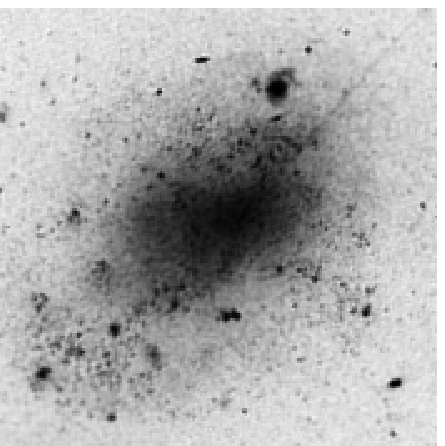}{IC3023}&
\FIG{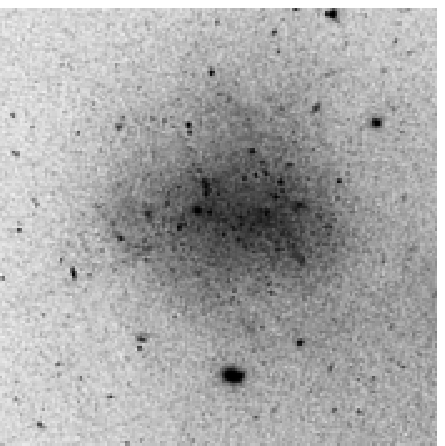}{GR34}&
\FIG{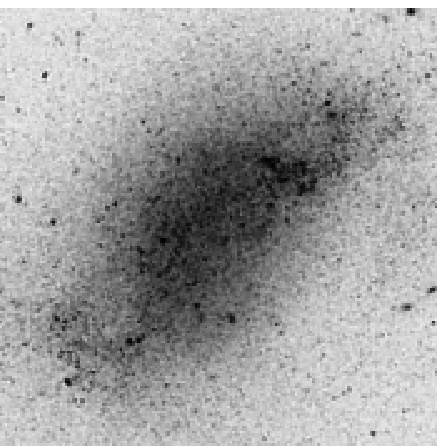}{U7512}\\
\FIG{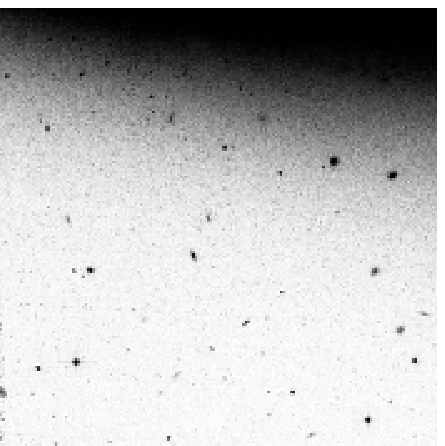}{N4517}&
\FIG{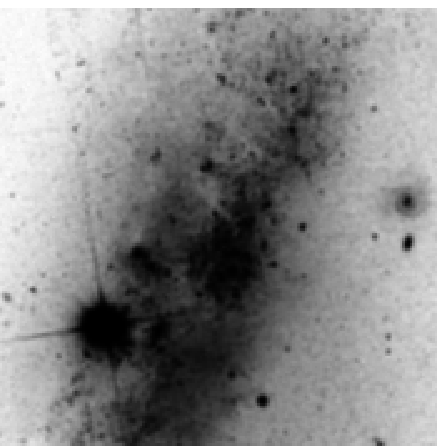}{IC3583}&
\FIG{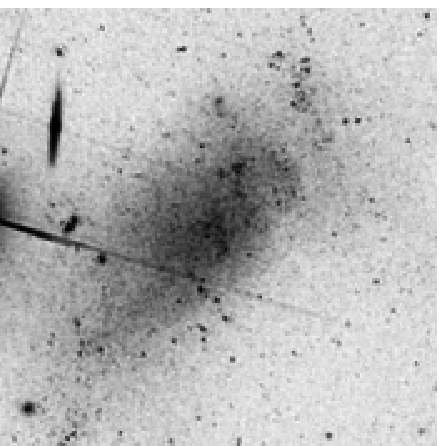}{KDG177}\\
\FIG{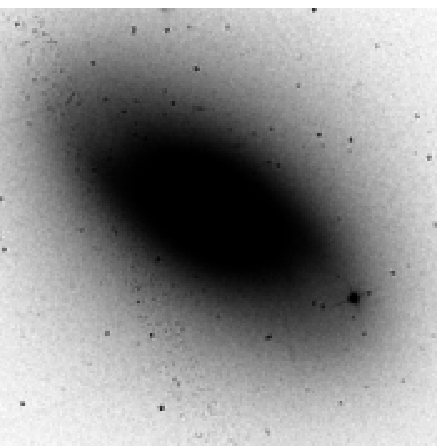}{N4600}&
\FIG{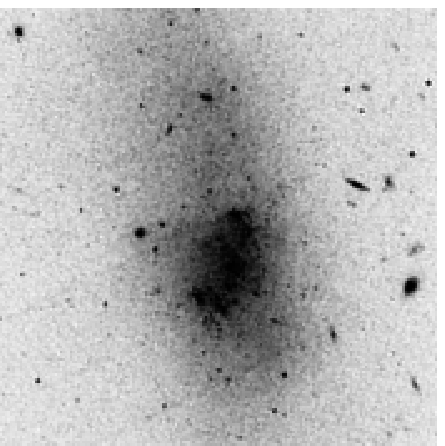}{VCC2037}&
\FIG{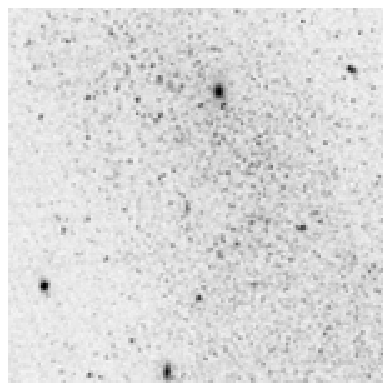}{KDG215}\\
\end{tabular}
\caption{Mosaic of enlarged ACS (F606W + F814W) images of
the nine galaxies. Field sizes are 1 arcminute on a side, North is up
and East is left.}
\end{figure}

\begin{figure}
\includegraphics[scale=1.0]{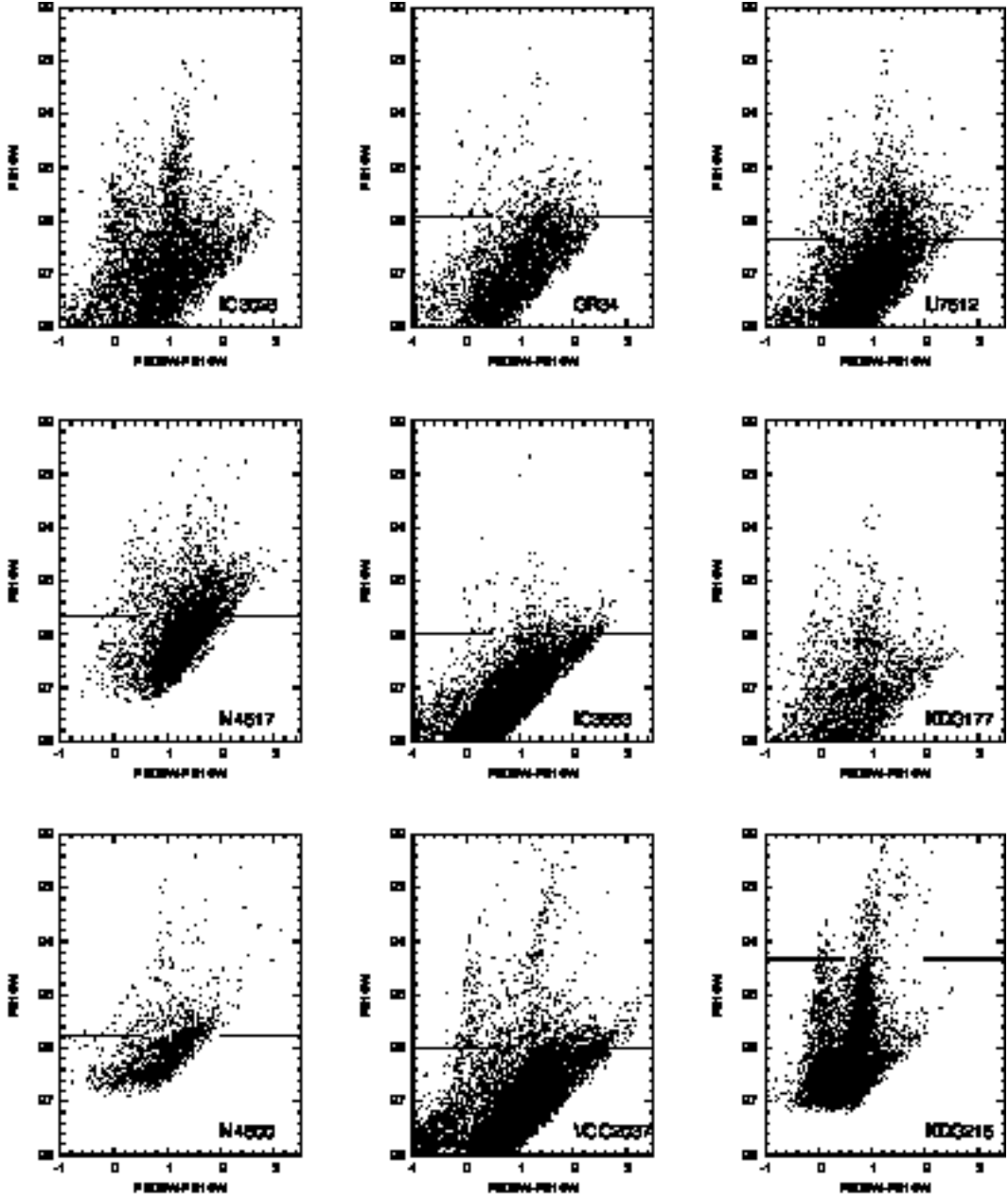}
\caption{Color--magnitude diagrams for nine target galaxies
from ACS observations. The broken horizontal lines mark the magnitudes of the TRGB.}
\end{figure}

\begin{figure}
\includegraphics[scale=0.75]{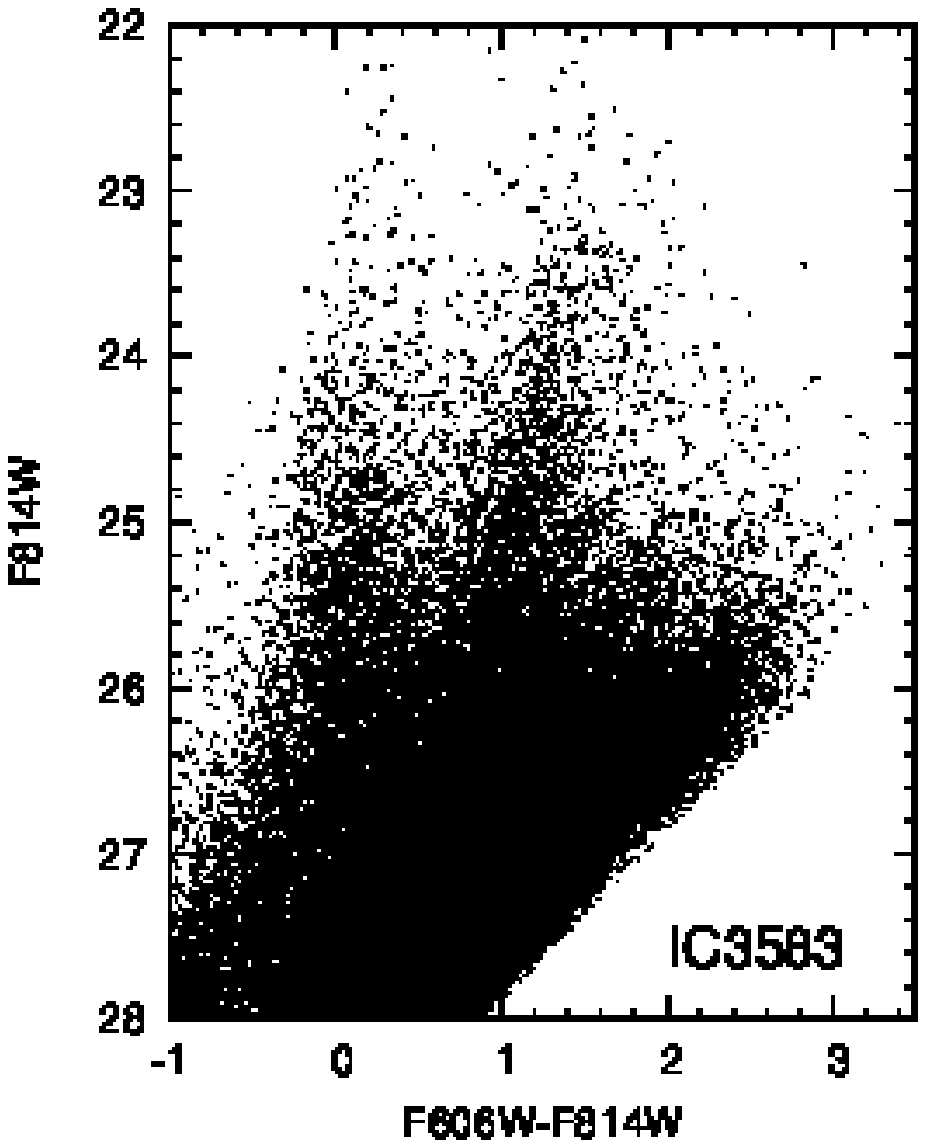}
\includegraphics[scale=0.75]{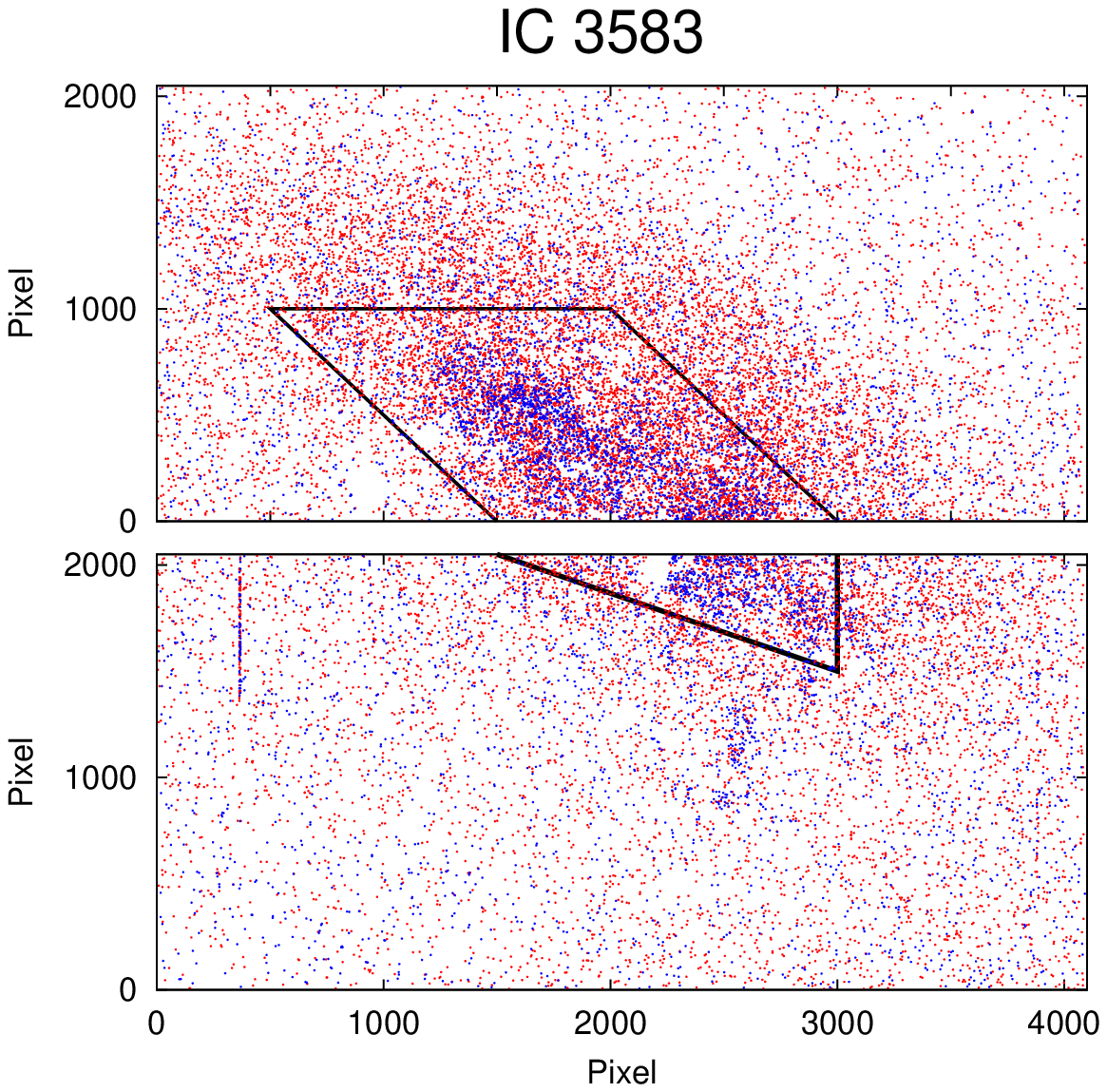}
\caption{Example of spatial clipping to reduce contamination from young populations
on the TRGB measurement.  The left panel shows the full field CMD for IC3583.
The right panel gives the positions of stars in the ACS field with stars with
F606W-F814W$>0.6$ in red and stars with F606W-F814W$<0.6$ in blue.  Blue stars are
concentrated toward the center.  Only stars outside
the exclusion box are included in the CMD for this galaxy shown in Fig. 4.}
\end{figure}

\begin{figure}[th!]
\includegraphics[width=\textwidth]{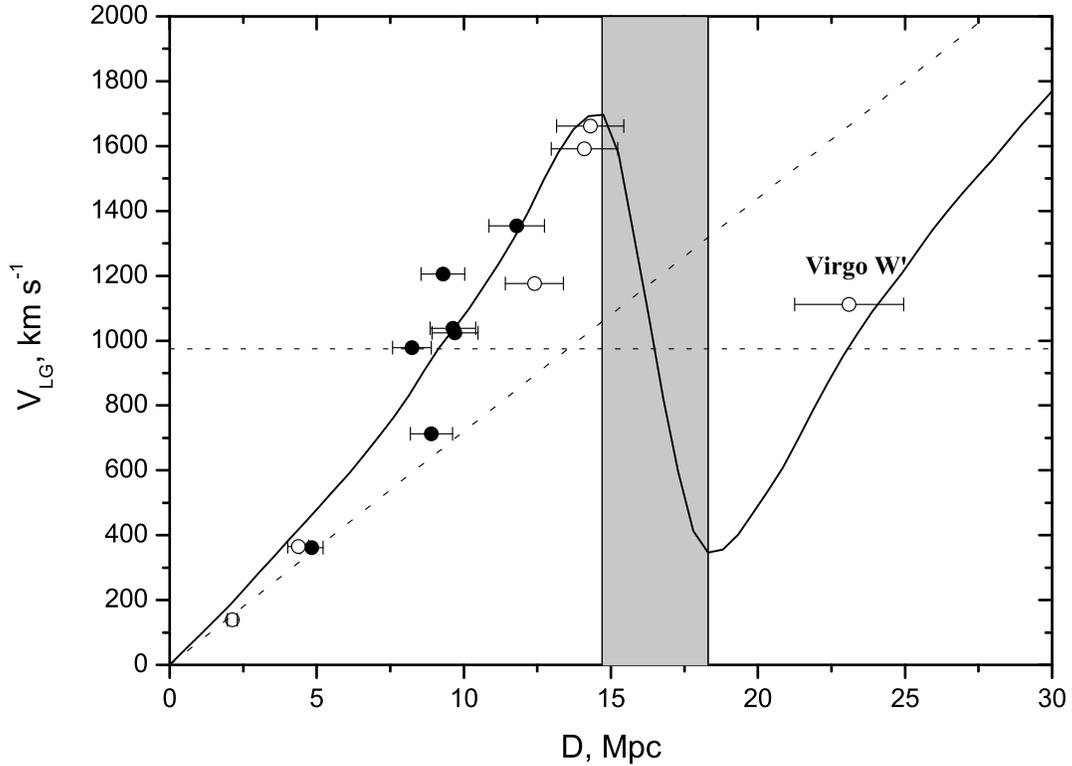}
\caption{Hubble flow in front of the Virgo cluster.  Filled symbols: galaxies
with new TRGB distance measures from HST observations (Table 1).  Open symbols:
galaxies with distances drawn from the literature (Table 2). The horizontal bars
indicate distance errors. The inclined dashed line marks the unperturbed Hubble flow.
The horizontal dashed line corresponds to the mean radial velocity of the Virgo cluster.
The grey vertical column denotes the zone of virial motions.}
 \end{figure}

\end{document}